\documentclass[iop]{emulateapj}
\usepackage{natbib}

\newcommand{\nltt}{NLTT~11748}
\def\Tref#1{Table~\ref{tab:#1}}
\def\Fref#1{Figure~\ref{fig:#1}}
\def\Sref#1{Section~\ref{sec:#1}}

\newcommand{\kms}{\ensuremath{{\rm km}\,{\rm s}^{-1}}}
\newcommand{\logg}{\ensuremath{\log(g)}}
\newcommand{\Teff}{\ensuremath{T_{\rm eff}}}
\newcommand{\til}{\raisebox{-0.7ex}{\ensuremath{\tilde{\;}}}}
\newcommand{\expnt}[2]{\ensuremath{#1 \times 10^{#2}}}   % scientific notation
\newcommand{\beq}{\begin{equation}}
\newcommand{\eeq}{\end{equation}}

\begin{document}

\title{Properties  of the Eclipsing Double-White Dwarf Binary \nltt}

\slugcomment{ApJ, in press}

\author{David L.~Kaplan\altaffilmark{1}, Thomas
  R.~Marsh\altaffilmark{2}, Arielle~N.~Walker\altaffilmark{1}, Lars
  Bildsten\altaffilmark{3,4}, 
  Madelon~C.~P.~Bours\altaffilmark{2}, Elm\'{e}~Breedt\altaffilmark{2},
  Chris~M.~Copperwheat\altaffilmark{5},
  Vik~S.~Dhillon\altaffilmark{6}, Steve B.~Howell\altaffilmark{7},
  Stuart~P.~Littlefair\altaffilmark{6}, Avi Shporer\altaffilmark{8},
  and Justin D.~R.~Steinfadt\altaffilmark{4}}
 
\altaffiltext{1}{Physics Department, University of Wisconsin-Milwaukee, Milwaukee WI 53211; kaplan@uwm.edu.}

\altaffiltext{2}{Department of Physics, University of Warwick, Coventry CV4 7AL, UK}

\altaffiltext{3}{Kavli Institute for Theoretical Physics and
  Department of Physics, Kohn Hall, University of California, Santa
  Barbara, CA 93106}

\altaffiltext{4}{Department of Physics, Broida Hall, University of
  California, Santa Barbara, CA 93106}

\altaffiltext{5}{Astrophysics Research Institute,
Liverpool John Moores University,
IC2, Liverpool Science Park,
146 Brownlow Hill,
Liverpool L3 5RF, UK}

\altaffiltext{6}{Department of Physics and Astronomy, University of
  Sheffield, Sheffield S3 7RH, UK}
\altaffiltext{7}{NASA Ames Research Center, Moffett Field, CA 94035, USA}
\altaffiltext{8}{California Institute of Technology, 1200 East California Boulevard, Pasadena, CA 91125, USA}

\begin{abstract}
We present high-quality ULTRACAM photometry of the eclipsing detached
double white dwarf binary \nltt.  This system consists of a
carbon/oxygen white dwarf and an extremely low mass (
$<0.2\,M_\odot$) helium-core white dwarf in a 5.6\,hr orbit.  To date
such extremely low-mass WDs, which can have thin, stably burning outer layers, have been
modeled via poorly constrained atmosphere and cooling calculations
where uncertainties in the detailed structure can strongly influence
the eventual fates of these systems when mass transfer begins.  With
precise (individual precision $\approx 1$\%), high-cadence ($\approx
2\,$s), multicolor photometry of multiple primary and secondary
eclipses spanning $>1.5\,$yr, we constrain the masses and radii of both
objects in the \nltt\ system to a statistical uncertainty of a few
percent. However, we find that overall uncertainty in the thickness of
the envelope of the secondary carbon/oxygen white dwarf leads to a larger
($\approx 13$\%) systematic uncertainty in the primary He WD's mass.  Over
the full range of possible envelope thicknesses, we find that our
primary mass (0.136--0.162$\,M_\odot$) and surface gravity
($\logg=6.32$--6.38; radii are 0.0423--0.0433$\,R_\odot$) constraints
do not agree with previous spectroscopic determinations.  We use
precise eclipse timing to detect the R{\o}mer delay at $7\sigma$
significance, providing an additional weak constraint on the masses
and limiting the eccentricity to $e\cos\omega=\expnt{(-4\pm5)}{-5}$.
Finally, we use multicolor data to constrain the secondary's
effective temperature ($7600\pm120\,$K) and cooling age
(1.6--1.7\,Gyr).
\end{abstract}


\section{Introduction}
Among the more interesting products of binary evolution are compact
binaries (periods less than 1 day) containing helium-core white dwarfs
(WDs).  These WDs are created from low-mass ($<2.0\,M_\odot$) stars
when stellar evolution is truncated by a binary companion before the
He core reaches the $\approx 0.48\, M_\odot$ needed for the helium
core flash.  Such WDs were first identified both as
companions to millisecond pulsars \citep{llfn95,vkbjj05,bvkkv06} and
other WDs \citep[e.g.,][]{bsl92,mdd95}, with large numbers
of double WD binaries discovered in recent years. In particular, the
Extremely Low Mass (ELM) survey \citep[][and references
  therein]{kbap+12,bkap+13} has discovered 10's of new He-core WDs in
the last few years, focusing on the objects with masses
$<0.2\,M_\odot$.

The compact binaries containing these WDs will inspiral due to
emission of gravitational radiation in less than a Hubble time; the
most compact of them will merge in $<1\,$Myr \citep{bkh+11}.  When
mass transfer begins detailed evolutionary and mass transfer
calculations \citep{mns04,davb+06,kbs12} will determine whether the
objects remain separate (typically resulting in an AM~CVn binary) or
merge (as a R~CrB star or possibly a Type~Ia supernova;
\citealt{it84,webbink84}).  Essential to determining the fates of
these systems (and hence making predictions for low-frequency
gravitational radiation and other end products) is an accurate knowledge
of their present properties: their masses determine the in-spiral
time and their radii and degrees of degeneracy help determine the
stability of mass transfer \citep{dbn05,davb+06,kbs12}.  This is
particularly interesting for the ELM WDs, as they are predicted to
possess stably burning H envelopes (with $ \sim 10^{-3}$--$10^{-2}\,
M_\odot$ of hydrogen) that keep them bright for Gyr
\citep{sarb02,pach07} and increase their radii compared with ``cold''
fully degenerate WDs by a factor of 2 or more.  This burning slows the
cooling behavior of these objects (it may not be monotonic for all
objects), and improving our understanding of ELM WD cooling would aid
in evolutionary models for millisecond pulsars and later stages of
mass transfer \citep[e.g.,][]{tlk12,avkk+12,kbvk+13}.

Few ELM WDs have had mass and radius measurements of any precision.
As most systems are single-line spectroscopic binaries \citep{kbs12}, precise masses
are difficult to obtain (although some pulsar systems are better;
\citealt{bvkkv06,avkk+12}).  Radii are even harder, typically relying
on poorly calibrated surface gravity measurements and cooling models
(as in \citealt{kbap+12}).  The eclipsing double WD binary
\object[NLTT 11748]{\nltt} \citep{sks+10} allowed for the first
geometric measurement of the radius of an ELM WD in the field
(cf.\ \object[PSR J1911-5958A]{PSR~J1911$-$5958A} in the globular
cluster NGC~6752; \citealt{bvkkv06}), finding $R\approx 0.04\,R_\odot$
for the $\approx 0.15\,M_{\odot}$ He WD, with new eclipsing systems
\citep{pmg+11,bkh+11,vtk+11} helping even more.  However, for \nltt\ the
original eclipse constraints from \citet{sks+10} were limited in their
precision.  As the system is a single-line binary, individual masses
were not known.  Further uncertainties came from limited photometric
precision and a low observational cadence, along with ignorance of
proper limb darkening for WDs of this surface gravity and temperature.

Here, we present new data and a new analysis of eclipse photometry for
\nltt\ that rectifies almost all of the previous limitations and gives
precise masses and radii that are largely model independent (at least
concerning models of the ELM WDs themselves), allowing for powerful new
constraints on the evolution and structure of ELM WDs.  \nltt\ was
identified by \citet{kv09} as a candidate ELM WD binary, containing a
helium-core WD with mass $\approx 0.15\,M_\odot$ presumably
orbiting with a more typical $0.6\,M_\odot$ carbon/oxygen (CO) WD
(note that the photometric primary is the lower-mass object, owing to
the inverted WD mass-radius relation).  While searching for
pulsations, \citet{sks+10} found periodic modulation in the light curve
of \nltt\ which they determined was due to primary (6\%) and secondary
(3\%) eclipses in a 5.6\,hr orbit, as confirmed by radial velocity
measurements (also see \citealt*{kvv10}).  The primary low-mass WD has
a low surface gravity ($\logg=6.18\pm0.15$ from \citealt{kvv10}
$\logg=6.54\pm0.05$ from \citealt{kapb+10b}) and an effective
temperature $\Teff=8580\pm50\,$K (\citealt{kvv10}, or $8690\pm140\,$K
from \citealt{kapb+10b}).  Constraints in this region of
$(\logg,\Teff)$ space are particularly valuable, as the behavior of
systems in this region is complex with a wide range of predicted ages
consistent changing over small ranges of mass, especially since it is
near the transition from systems that show CNO flashes and those that
do not \citep*{ambc13}.  

Our new data consist of high-cadence (2.5\,s compared with 30\,s
previously) high-precision photometry in multiple simultaneous
filters, which we combine with improved modeling and knowledge of
limb-darkening coefficients \citep{gskb13}.  We outline the new
observations in \Sref{obs}.  The majority of the new analysis is
described in \Sref{fit}, with the results in \Sref{results}.  Finally,
we make some additional physical inferences and discuss our results
in \Sref{discuss}.

\section{Observations and Reduction}
\label{sec:obs}
\begin{deluxetable*}{l c c c c c c c}
\tablewidth{0pt}
\tabletypesize{\footnotesize}
\tablecaption{Log of ULTRACAM observations and eclipse times\label{tab:log}}
\tablehead{
\colhead{Date} & \colhead{Eclipse Time} & \colhead{Telescope} &
\colhead{Eclipse} & \colhead{Filters\tablenotemark{a}} &
\colhead{Exposures\tablenotemark{a}} &
\colhead{Num.\ Stars\tablenotemark{a}}  & \colhead{Precisions\tablenotemark{a}}\\
 & \colhead{(MBJD)} & & & & \colhead{(s)} & & \colhead{(\%)}
}
\startdata
2010 Nov 12\ldots & $55512.182179(17)$ & NTT & secondary & $u^\prime
g^\prime i^\prime$ & 7.69,2.55,2.55 & 4,5,3 & 2.9,1.0,1.3\\
2010 Nov 15\ldots & $55515.120443(21)$ & NTT & primary & $u^\prime
g^\prime i^\prime$ & 7.69,2.55,2.55 & 4,5,4 & 6.4,1.8,2.2\\
2010 Nov 15\ldots & $55515.237910(23)$ & NTT & secondary & $u^\prime
g^\prime i^\prime$ & 7.69,2.55,2.55 & 4,5,3 & 3.6,1.2,1.7\\
2010 Nov 25\ldots & $55525.228090(18)$ & NTT & primary & $u^\prime
g^\prime i^\prime$ & 7.69,2.55,2.55 & 4,5,3 & 4.2,1.5,1.6\\
2010 Nov 26\ldots & $55526.168300(13)$ & NTT & primary & $u^\prime
g^\prime r^\prime$ & 5.89,1.95,1.95 & 4,5,3 & 3.5,1.3,1.6\\
2010 Nov 26\ldots & $55526.285767(29)$ & NTT & secondary & $u^\prime
g^\prime r^\prime$ & 5.89,1.95,1.95 & 4,5,2 &  6.6,1.9,2.5\\
2010 Nov 27\ldots & $55527.108548(12)$ & NTT & primary & $u^\prime
g^\prime i^\prime$ & 5.89,1.95,1.95 & 4,5,4 & 3.4,1.1,1.5\\
2010 Nov 27\ldots & $55527.226026(20)$ & NTT & secondary & $u^\prime
g^\prime i^\prime$ & 5.46,1.35,1.35 & 4,5,4 & 3.8,1.5,1.8\\
2010 Nov 28\ldots & $55528.166253(16)$ & NTT & secondary & $u^\prime
g^\prime r^\prime$ & 5.89,1.95,1.95 & 4,5,5 & 3.3,1.1,1.3\\
2010 Nov 29\ldots & $55529.106526(19)$ & NTT & secondary & $u^\prime
g^\prime i^\prime$ & 5.46,1.35,1.35 & 4,4,4 & 4.1,1.4,2.0\\
2010 Dec 02\ldots & $55532.162334(16)$ & NTT & secondary & $u^\prime g^\prime i^\prime$ & 7.69,2.55,2.55 & 4,4,4 &2.6,0.9,1.3\\
2010 Dec 10\ldots & $55540.154399(16)$ & NTT & secondary & $u^\prime g^\prime i^\prime$ & 7.69,2.55,2.55 & 4,4,4 &2.7,0.9,1.3\\
2010 Dec 15\ldots & $55545.208237(16)$ & NTT & primary & $u^\prime g^\prime i^\prime$ & 7.48,2.48,2.48 & 4,4,4 &3.7,1.3,1.7\\
2010 Dec 16\ldots & $55546.148475(18)$ & NTT & primary & $u^\prime g^\prime i^\prime$ & 7.48,2.48,2.48 & 4,4,4 &4.7,1.6,2.0\\
2010 Dec 17\ldots & $55547.088715(23)$ & NTT & primary & $u^\prime g^\prime i^\prime$ & 7.48,2.48,2.48 & 4,4,4 &5.1,1.8,2.1\\
2010 Dec 18\ldots & $55548.146459(36)$ & NTT & secondary & $u^\prime g^\prime i^\prime$ & 7.48,2.48,2.48 & 4,4,4 &7.0,2.6,3.4\\
2012 Jan 17\ldots & $55943.870809(09)$ & WHT & primary & $u^\prime
g^\prime r^\prime$ & 2.48,2.48,2.48 & 4,5,3 & 4.1,0.7,0.8\\
2012 Jan 17\ldots & $55943.988273(13)$ & WHT & secondary & $u^\prime
g^\prime r^\prime$ & 4.99,2.48,2.48 & 4,5,4& 2.9,0.8,0.9\\
2012 Jan 18\ldots & $55944.928546(22)$ & WHT & secondary & $u^\prime
g^\prime r^\prime$ & 4.99,2.48,2.48 & 4,5,3 &4.1,1.2,1.5\\
2012 Jan 19\ldots & $55945.046118(28)$ & WHT & primary & $u^\prime
g^\prime r^\prime$ & 4.98,2.48,2.48 & 4,5,3 & 5.3,1.7,2.0\\
2012 Jan 19\ldots & $55945.868766(42)$ & WHT & secondary & $u^\prime
g^\prime$ & 5.99,2.98,\nodata & 4,5,\nodata & 3.7,1.3,\nodata\\
2012 Jan 21\ldots & $55947.866825(09)$ & WHT & primary & $u^\prime
g^\prime r^\prime$ & 3.46,1.72,1.72 & 4,5,2 & 2.9,0.9,1.0\\
2012 Jan 22\ldots & $55948.924551(12)$ & WHT & secondary & $u^\prime
g^\prime r^\prime$ & 3.99,1.98,1.98 & 4,5,4 & 2.9,0.8,0.9\\
2012 Jan 23\ldots & $55949.042137(10)$ & WHT & primary & $u^\prime
g^\prime r^\prime$ & 3.99,1.98,1.98 & 4,5,4 & 4.0,0.9,1.1\\
2012 Sep 01\ldots & 56171.174286(11) & WHT & primary & $u^\prime g^\prime r^\prime$ &
5.56,1.84,1.84 & 4,4,4 & 3.8,1.6,1.1\\
2012 Sep 04\ldots & 56174.230072(11) & WHT & primary & $u^\prime g^\prime r^\prime$ &
6.48,2.14,2.14 & 5,4,4 & 2.5,1.0,1.0 \\
2012 Sep 10\ldots & 56180.106572(12) & WHT & primary & $u^\prime g^\prime r^\prime$ &
5.00,2.48,2.48 & 4,4,4 & 2.8,2.1,0.9 \\
\enddata
\tablenotetext{a}{We give the three filters used along with the
  corresponding exposure times, number of reference stars used, and typical fractional precisions on a
  single measurement of \nltt.}
\end{deluxetable*}

\subsection{ULTRACAM Observations}
We observed \nltt\ with ULTRACAM \citep{dms+07} over 27 eclipses
during 2010 and 2012, as summarized in Table~\ref{tab:log}.  ULTRACAM
provides simultaneous fast photometry through 3 filters with
negligible dead time.  During 2010, it was mounted on the 3.5\,m New
Technology Telescope (NTT) at La Silla Observatory, Chile.  We used
the $u^\prime$ and $g^\prime$ filters, along with either $r^\prime$ or
$i^\prime$. The integration times were chosen based on the conditions,
but were typically 1--2\,s for the redder filters and 5--8\,s for the
$u^\prime$ filter.  During 2012 ULTRACAM was mounted on the 4.2\,m
William Herschel Telescope at the Observatorio del Roque de los
Muchachos on the island of La Palma.  Here, we used only the $u^\prime
g^\prime r^\prime$ filters, although for one observation we discarded
the $r^\prime$ data as they were corrupted.  Exposure times were 2--3\,s
for the redder filters and 3--5\,s for the $u^\prime$ filter, taking
advantage of better conditions, a larger mirror, and a lower airmass
toward this northern target.  The total observing time for each
eclipse was typically less than 40\,minutes.

The data were reduced using custom software.  We first determined bias
and flatfield images appropriate for every observation.  Then, we
measured aperture photometry for \nltt\ and up to 6 reference sources
that were typically somewhat brighter than \nltt\ itself.  The
aperture was sized according to the mean seeing for the observation,
but was held fixed for a single observation and a single filter.
Finally, a weighted mean of the reference star magnitudes (some of
which were removed because of saturation and some of which were
removed because of low signal-to-noise, especially in the $u^\prime$
band) was subtracted from the measurements for \nltt.  These detrended
data were the final relative photometry that we used in all subsequent
analysis.  Given the short duration of the eclipse ($\approx 3\,$minutes),
any remaining variations in the relative photometry due to
transparency or airmass changes could be ignored; we did not attempt
to model the out-of-eclipse data \citep[cf.][]{sks+10b}.

\subsection{Near-Infrared Observations}
In addition to the ULTRACAM observations, we observed a few eclipses
using the Gemini Near-Infrared Imager (NIRI; \citealt{hji+03}) on the
8\,m Gemini-North telescope under program GN-2010B-Q-54.  Given the low
signal-to-noise and relatively long cadence, the primary eclipse data
were not particularly useful in constraining the properties of the
system.  Instead, we
concentrate on the secondary eclipse observations, where the
additional wavelength coverage is helpful in constraining the
effective temperature of the secondary (\Sref{temp}).  The data were
taken on 2010~November~21 and 2010~December~15 using the $J$-band
filter.  The total duration of the observations were 37 and 31 minutes
around a secondary eclipse, as predicted from our initial ephemeris.
Successive exposures happened roughly every 25\,s, of which 20\,s were
actually accumulating data, so our overhead was  about 20\%.

To reduce the data, we used the \texttt{nprepare} task in
\texttt{IRAF}, which adds various meta-data to the FITS files.  We
then corrected the data for nonlinearities\footnote{Using the
  \texttt{nirlin.py} script from
  \url{http://staff.gemini.edu/{\til}astephens/niri/nirlin/nirlin.py}.}
and applied flatfields computed using the \texttt{niflat} task, which
compared dome-flat exposures taken with the lamp on and off to obtain
the true flatfield.  We used our own routines to perform point spread
function photometry with a \citet{moffat69} function.  This function
was held constant for each observation and was fit to bright
reference stars.  Finally, we subtracted the mean of two bright
reference stars to de-trend the data.

\begin{figure*}[t]
%\plotone{../R5_650_1.000_flux.eps}
\plotone{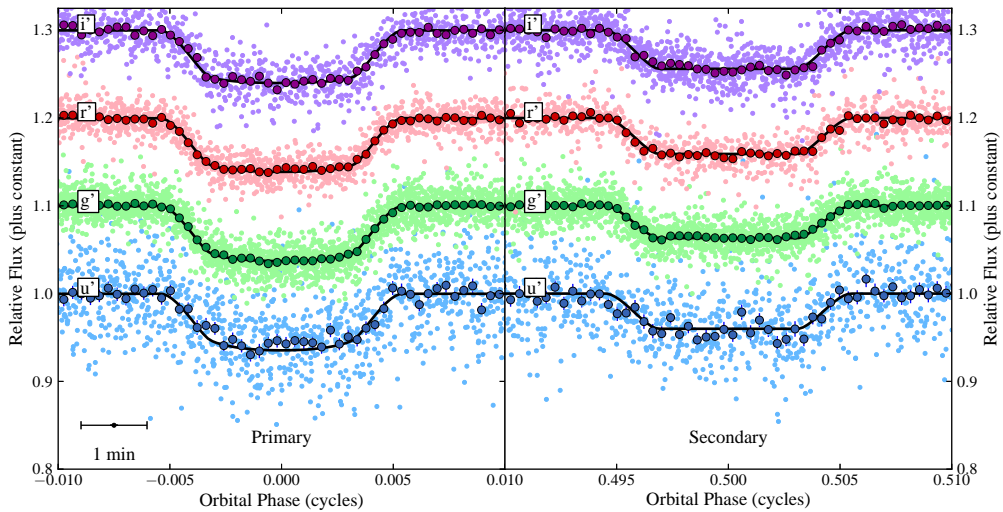}
\caption{Normalized primary (left) and secondary (right) eclipses of \nltt, as
  measured with ULTRACAM. The raw data are the points, while binned data are the
  circles with error bars, and  the
  best-fit models are the solid lines.  The different filters are
  labeled.  A 1\,minute interval is indicated by the scale bar at the
  lower left.  Data from 2010 and 2012 have been combined. }
\label{fig:eclipses}
\end{figure*}

\section{Eclipse Fitting}
\label{sec:fit}
To start, we determined rough eclipse times and shapes by fitting a
simple model to the data, using a square eclipse for the secondary
eclipses and a linear limb-darkening law for the primary eclipses.
These results were only used as a starting point for the later
analysis, but the times were correct to $\pm10\,$s.  We then fit the
photometry data, as summarized below.  The fitting used only the
ULTRACAM data; the NIRI data were added later to constrain the
secondary temperature.

Our main eclipse fitting used a Markov-Chain Monte Carlo (MCMC)
fitter, based on a \texttt{Python} implementation\footnote{See
  \url{http://dan.iel.fm/emcee/}.}  \citep{fmhlg13} of the
affine-invariant ensemble sampler \citep{gw10}.  
We parameterized the light curve according to
\begin{itemize}
\item Mass and radius of the primary (low-mass) WD, $M_1$ and
  $R_1$
\item Orbital inclination $i$
\item Radial velocity amplitude of the primary, $K_1$
\item Mean period $P_B$, reference time $t_0$, and time delay $\Delta t$
\item Temperatures of the primary $T_1$ and secondary $T_2$
\end{itemize}
for 9 total parameters.  These were further constrained by priors
based on spectroscopy, with $K_1=273.4\pm0.5\,\kms$ and
$T_1=8690\pm140\,$K \citep{sks+10,kapb+10b}.  We assumed a
strictly periodic ephemeris (with no spin down; see \Sref{ephem}) that
includes a possible time delay between the primary and secondary
eclipses \citep{kaplan10}.  

Our limb-darkening law used 4-parameter \citep{claret00}
limb-darkening coefficients, as determined by \citet{gskb13} for a
range of gravities and effective temperatures.  We interpolated the
limb-darkening parameters for the primary's temperature $T_1$,
although we used several fixed values of \logg\ (6.25, 6.50, and 6.75)
instead of the value for each fit.  This is done both to avoid
numerical difficulties in two-dimensional (2D) interpolation over a coarse
grid and to avoid biasing the fitted \logg\ by anything other than the
light-curve shape (unlike the temperature, the different spectroscopic
determinations of the surface gravity are significantly discrepant).
We found that variations in  \logg\ used for the limb-darkening
parameters changed the fit results by $<1\,\sigma$.

With values for $M_1$, $i$, $P_B$, and $K_1$, the mass of the
secondary (high-mass) WD, $M_2$, is then determined (and so is the
mass ratio $q\equiv M_1/M_2$, as well as $K_2=q K_1$).  The final
parameter is the radius of the secondary WD $R_2$.  However, as this
is a more or less normal CO WD that is not tidally distorted
($(M_1/M_2)(R_2/a)^3\approx 10^{-7}$), we used a mass-radius relation
appropriate for WDs in this mass range
\citep*{fbb01,blr01}\footnote{See
  \url{http://www.astro.umontreal.ca/{\til}bergeron/CoolingModels/}.},
interpolating linearly.  The complication that this introduced is that
WDs with finite temperatures do have radii slightly larger
than the nominal zero-temperature model and that this excess depends
on the thickness of their hydrogen envelopes.  For the mass and
temperature range considered here, this excess is typically $r_2\equiv
R_2/R_2(T_2=0)=1.02$--1.06.  In what follows, we treat $r_2$ as a free
parameter and give our results in terms of $r_2$, with a detailed
discussion of the influence of $r_2$ on the other parameters in
\Sref{r2}.

Overall, we had 22,574 photometry measurements within $\pm 400\,$s of
the eclipses (shown in \Fref{eclipses}), which
we corrected to the solar system barycenter using a custom extension
to the \texttt{TEMPO2}\footnote{See
  \url{http://www.atnf.csiro.au/research/pulsar/tempo2/}.} pulsar
timing package \citep*{hem06}.  The eclipses themselves were modeled
with the routines of \citet[][also see \citealt{ma02}]{agol02}, which
accounts for intrabinary lensing \citep{maeder73,marsh01}.

We started 200 MCMC ``walkers'', where each walker executes an
independent path through the parameter space.  The walkers were
initialized from normally distributed random variables, with each
variable taken from the nominal values determined previously
\citep{sks+10,kapb+10b} with generous uncertainties.  In the end, we
increased the uncertainties on the initial conditions and it did not
change the resulting parameter distributions.  Each walker was allowed
500 iterations to ``burn in'', after which its memory of the sampled
parameter space was deleted.  Finally, each walker iterated for a
further 5000 cycles, giving $200\times5000=1,000,000$ samples.  However,
not all of these are independent: we measured an auto-correlation
length of about 100 samples from the resulting distributions, so we
thinned the parameters by taking every 91 samples (we wanted a number
near our measured auto-correlation but that was not commensurate with
the number of walkers).

\subsection{The Influence of the Secondary's Envelope Thickness on the Measurement}
\label{sec:r2}
As discussed above, our one significant assumption (which was also
made in \citealt{sks+10}) is that the secondary star follows the
mass-radius relation for a CO WD.  This seems reasonable, given
inferences from observations \citep{kvv10,kapb+10b} and from
evolutionary theory.  However, with the high precision of the current
dataset we must examine the choice of mass-radius relation closely.
In particular, the zero-temperature model used in \citet{sks+10} is no
longer sufficient.  For effective temperatures near 7500\,K and masses
near 0.7\,$M_\odot$, finite-temperature models are larger than the
zero-temperature models by roughly 2\% (thin envelopes, taken to be
$10^{-10}$ of the star's mass) to 6\% (thick envelopes, taken to be
$10^{-4}$ of the star's mass) and moreover the excess depends on mass
\citep{fbb01,blr01}.  This excess is similar to what we observe in a
limited series of models computed using \texttt{Modules for
  Experiments in Stellar Astrophysics} (\citealt{pbd+11,pca+13}).  The
envelope thickness can be constrained directly through astroseismology
of pulsating WDs (ZZ Ceti stars), with most sources having thick
envelope (fractional masses of $10^{-6}$ or above), but extending down
such that roughly 10\% of the sources have thin envelopes (fractional
masses of $10^{-7}$ or below); there is a peak at thicker envelopes,
but there is a broad distribution (\citealt{rca+12}; consistent with
the findings of \citealt{tb08}).

\begin{figure*}[t]
%\plotone{../R5_650_1.000.eps}
\plotone{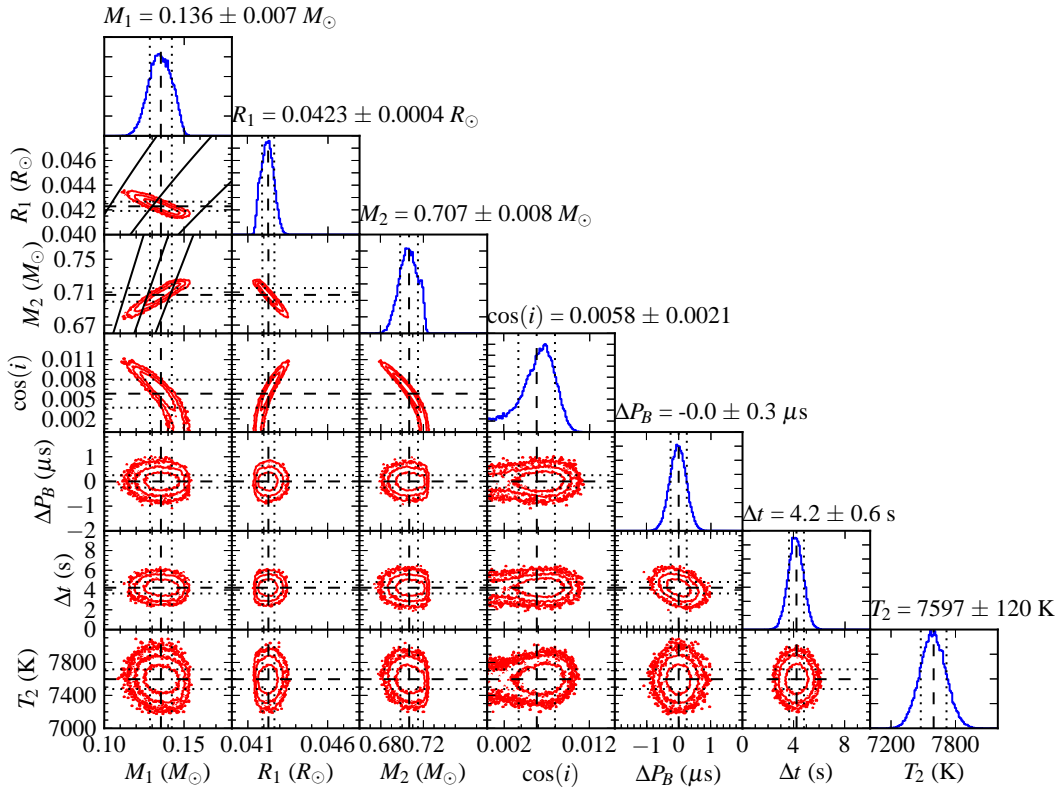}
\caption{Joint confidence contours on the parameters from the 
  fit of \nltt, assuming $r_2=1.00$.  We show 68\%, 95\%, and 99.7\%
  contours on WD distributions that have been
  marginalized from the 8D original distribution (we do
  not plot distributions for the reference time $t_0$, as it is of
  little physical interest).  $\Delta P_B$ is the offset of $P_B$ with
  respect to its mean (\Tref{fullfitr2}).  We also show the
  1D distributions for each parameter.  In all of the
  plots, the black dashed lines show the means and the black dotted
  lines show the $\pm1\,\sigma$ limits.  Finally, in the plot of $M_2$
  vs.\ $M_1$ we show solid lines corresponding to the mass ratios
  $q=0.16,0.18$, and0.20 (which can map to constraints from $\Delta t$),
  while in the plot of $R_1$ vs.\ $M_1$ we show solid lines
  corresponding to $\logg=6.2,6.3$, and 6.4.}
\label{fig:confidence}
\end{figure*}

We use the parameter $r_2$ to explore the envelope thickness.  Values
near $1.02$ correspond to thin envelopes (with some slight mass
dependence), while values near 1.06 correspond to thick envelopes.
Changing $r_2$ over the range of values discussed above leads to
changes in the best-fit physical parameters $M_1$, $M_2$, $R_1$, and
$R_2$.  In particular, $M_1$ was surprisingly sensitive to
$r_2$. Determining a ``correct'' value for $r_2$ is beyond the scope
of this work, but we can understand how the physical parameters scale
with $r_2$ in a reasonably simple manner.  Since we know the period
accurately and can also say that $\sin i\approx 1$ (deeply
eclipsing), we know that $M_1+M_2\propto a^3$ (where  $a$ is the semi-major
axis) from Kepler's third law.
We also know $K_1$, which is  the orbital speed of the
primary:
\beq
K_1 \approx \frac{2\pi a}{P_B} \frac{M_2}{M_1+M_2},
\eeq
which  allows us to constrain $a M_2 \propto M_1+M_2$.
Combining these gives $M_2\propto a^2$.  We can further parameterize
the mass-radius relation of the secondary:
\beq
R_2 \propto r_2 {M_2^\beta}.
\label{eqn:r2}
\eeq
The duration of the eclipse fixes $R_1/a$, while the duration of
ingress/egress fixes $R_2/a$ \citep[e.g.,][]{winn11}, so we can also say
$a\propto R_2$ or $a\propto r_2 M_2^\beta$.  Combining this with $M_2
\propto a^2$ gives $M_2 \propto r_2^{2/(1-2\beta)}$.  Near $M_2\approx
0.7\,M_\odot$, the mass--radius relation has a slope $\beta\approx
-0.78$ (compare with the traditional $\beta=-1/3$ for lower masses), so
$M_2\propto r_2^{0.78}$.  Since $a\propto M_2^{1/2}$ and both $R_1$
and $R_2$ are $\propto a$, $R_1\propto r_2^{1/(1-2\beta)}\propto
r_2^{0.39}$ and the same for $R_2$ (at a fixed $M_2$, $R_2\propto r_2$,
as given in Eqn.~\ref{eqn:r2}; however, the result here is for the
best-fit value of $R_2$, which can result in changes of the other
parameters including $M_2$).

To understand how $M_1$ changes with $r_2$, we use the Keplerian mass
function, which fixes $M_2^3 \propto (M_1+M_2)^2$.  If we take the
logarithmic derivative of this, we find:
\beq
\frac{d\log M_1}{M_1} = \alpha \frac{d\log M_2}{M_2}
\eeq
with $\alpha=3/2 + M_2/2M_1$.  Since our mass ratio $q$ is roughly
$0.2$ (\Tref{fullfitr2}), we find $\alpha\approx 4.0$.  From this, we
find that $M_1 \propto r_2^{2\alpha/(1-2\beta)}\propto r_2^{3.13}$,
which is much steeper than the other dependencies.  These relations
are borne out by MCMC results (\Tref{fullfitr2}, \Fref{confidence},
and \Fref{mr}).

\subsection{Results}
\label{sec:results}

The model fit the data well, with a minimum $\chi^2$ of $22978.8$ for
22,566 degrees of freedom\footnote{In all of our fitting, we make use
  of the $\chi^2$ statistic.  This assumes that individual data points
  are independent of each other; on the other hand, correlated errors
  can become significant for very precise photometry and can alter the nature of
  parameter and uncertainty estimation \citep[see][]{cw09}.  To test
  this, we examined the out-of-eclipse data for any correlation between
  subsequent data points.  We found autocorrelation lengths of 0--2
  samples, with a mean of 0.85.  This was very similar to the
  distribution of autocorrelation lengths estimated from sets of 100
  uncorrelated random numbers drawn from ${\cal N}(1,0.003)$ (similar
  in length and properties to our data), so we conclude that
  deviations from an autocorrelation length of 0 are consistent with
  the finite sample sizes that we used and that the data are
  consistent with being independent.}.  We show 1D and 2D marginalized confidence contours in \Fref{confidence}
and the best-fit light curves in 
\Fref{eclipses}. The results are given in
\Tref{fullfitr2}.  A linear ephemeris gives a satisfactory fit to the
data, although we find a significantly non-zero value for $\Delta t$,
which we discuss below.

For NIRI, the quality of the data is modest, with typical
uncertainties of $\pm 0.03$\,mag and a cadence of 25\,s. Given the
quality of the ULTRACAM results,  fitting the NIRI data with all parameters
free would not add to the results.  Instead, we kept the physical
parameters fixed at their best-fit values from \Tref{fullfitr2} and
 only fit for the eclipse depth at $1.25\,\mu$m.  

The results are shown in \Fref{niri}.  Despite the modest quality of
the data the fit is good, with $\chi^2=160.7$ for 161
dof.  We find a depth $d_2(1.25\,\mu{\rm m})$ of
$4.2\%\pm0.4$\%, corresponding to a $J$-band flux ratio
$d_2(1.25\,\mu{\rm m})/(1-d_2(1.25\,\mu{\rm m}))$ of $4.4\%\pm0.4$\% (see \Sref{temp}).
This value is slightly off from the predictions based on our fit to
the ULTRACAM data (\Fref{depth}), although by less than 2$\sigma$.

\begin{figure}[t]
%\plotone{../nltt11748_mr.eps}
\plotone{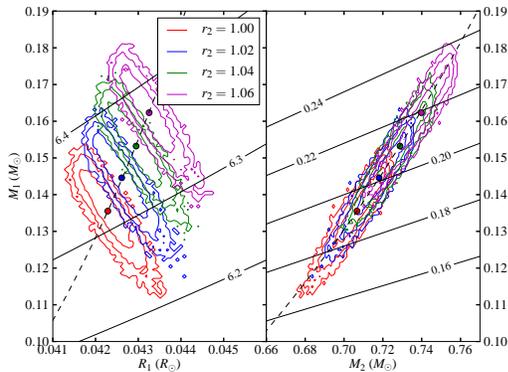}
\caption{Mass and radius constraints as a function of $r_2$, showing
  $r_2=1.00$ (red), $1.02$ (blue), 1.04 (green), and 1.06 (purple).
  In the left panel, we plot $M_1$ vs.\ $R_1$, while in the right panel
  we plot $M_1$ vs.\ $M_2$.  In both panels we plot the best-fit
  points (filled circles), along with the expected variations
  according to \Sref{r2} (dashed lines).  Additionally we plot contours of constant
  surface gravity $\logg=6.2,6.3$, and 6.4 (left) and mass ratio
  $q=0.16,0.18,0.20,0.22$, and 0.24 (right) in black.  }
\label{fig:mr}
\end{figure}

To derive the eclipse times in \Tref{log}, we used the results of the
full eclipse fit but fit the model (with the shape parameters held
fixed) to each observation individually.

\section{Discussion}
\label{sec:discuss}
The analyses presented in \Sref{fit} show precise determinations of
the masses and radii of the WDs in the \nltt\ binary.  We have one
remaining free parameter, which is the size of the radius excess of
the CO WD $r_2$, related to the size of its hydrogen envelope.
Moreover, we have shown that the eclipse data are consistent with a
linear ephemeris, although there is a systematic shift between the
primary and secondary eclipses.  Separately, the variation of the
secondary eclipse depth with wavelength allows for accurate
determination of the temperature of the CO WD, which then determines
its age through well-studied CO WD cooling curves.  Below, we discuss
additional constraints on the masses, radii, and ages of the
components determined by consideration of the eclipse times, secondary
temperature, and distance (determined by astrometry).

\begin{figure}[t]
%\plotone{../nltt11748_secondary_J.eps}
\plotone{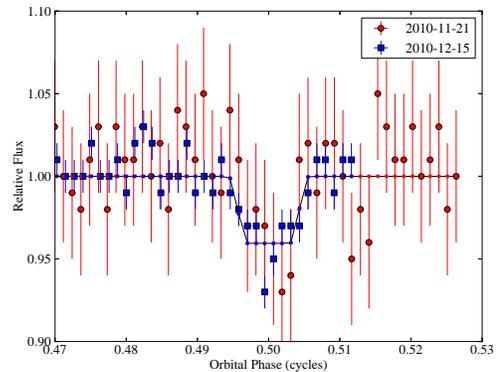}
\caption{Secondary eclipses of \nltt\ observed with Gemini/NIRI.  The
  two observations are the circles/squares, as labeled.  The solid
  curve is the best-fit model, with the points representing the model
  integrated over the 20\,s exposures for each set of observations.}
\label{fig:niri}
\end{figure}

\begin{figure}
%\plotone{../nltt11748_secondary_depths.eps}
\plotone{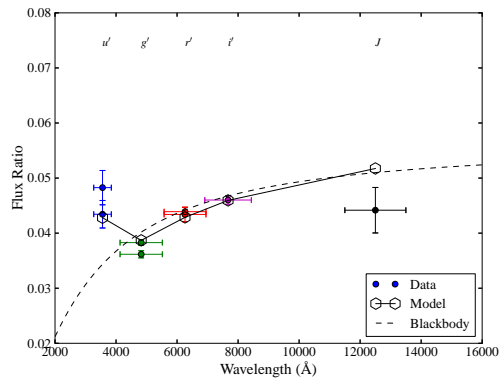}
\caption{Secondary-to-primary flux ratio as a function of wavelength
  for the secondary eclipses of \nltt.  We plot data from 2010 and
  2012 together, along with our best-fit model (open symbols).  The
  different bands are labeled.  The model is derived from
  \citet{tbg11} synthetic photometry.  We also show the corresponding
  flux ratio determined from blackbodies (which had been used
  previously), which does not match the $u^\prime$ data at all.}
\label{fig:depth}
\end{figure}

\subsection{Ephemeris, R{\o}mer Delay, and Mass Ratio Constraints}

\label{sec:ephem}
Using the measured eclipse times from Table~\ref{tab:log}, along with
those reported by \citet{sks+10}, we computed a linear ephemeris for
\nltt\ with a constant frequency $f_B=1/P_B$.  The residuals (\Fref{dt})
are consistent with being flat and with the results from the full
eclipse fitting (\Tref{fullfitr2}), showing   no indication
of  orbital changes.  However, we do find a systematic offset
between the times of the primary and secondary eclipses, as predicted
in \citet{kaplan10}.  The secondary eclipses arrive earlier on average,
by $\Delta t=4.1\pm0.5$\,s  (after correcting for this, the rms
residual for the new data is 1.7\,s and the overall $\chi^2$ for the
ephemeris data is 38.9 with 29 dof), which is consistent with $\Delta
t=4.2\pm0.6$\,s
inferred from the full eclipse fitting (\Tref{fullfitr2}).   The sign
is  correct  for a delay caused
by the light-travel delay across the system (the R{\o}mer delay) when
the more massive object is smaller,
 the magnitude of which is \citep{kaplan10}: 
\beq \Delta t_{\rm
  LT} = \frac{P_B K_1}{\pi c}\left(1-q\right).  
\eeq 
If we use $\Delta
t=4.2\pm0.6$\,s (from the full eclipse fitting in \Tref{fullfitr2}), then we infer $q_{\rm LT}=0.29\pm0.10$.  This is fully
consistent with our fitted values for $q$ (\Tref{fullfitr2};
$q=0.192\pm0.008$ for $r_2=1.00$).  However, it is also possible that
some time delay is caused by a finite eccentricity of the orbit
\citep{kaplan10,winn11}, with 
\beq 
\Delta t_{e}=\frac{2P_Be}{\pi}\cos \omega,
\eeq 
where $e$ is the eccentricity and $\omega$ is the argument
of periastron\footnote{Note that the expression from \citet{kaplan10}
  is missing a factor of two, as pointed out by \citet*{bwl12}.}.
However, the R{\o}mer delay \textit{must} be present in the eclipse
timing with a magnitude $(4.76\pm0.05)r_2^{2.6}\,$s, based on our mass
determination.  Therefore, instead of using the R{\o}mer delay to
constrain the masses, we can use it to constrain the eccentricity.
Doing this gives $e\cos\omega=\expnt{(-4\pm5)}{-5}$ (consistent with
a circular orbit).  This would then be one of the strongest constraints on the
eccentricity of any system without a pulsar, as long as the value of
$\omega$ is not particularly close to $\pi/2$ (or $3\pi/2$).  This may be testable
with long-term monitoring of \nltt, as 
relativistic apsidal precession \citep{bt76} should be $\dot \omega
\approx 2\deg\,{\rm yr}^{-1}$ (for a nominal $e=10^{-3}$).  As long as any tidal
precession is on a longer timescale, the change in $\omega$ could
separate the R{\o}mer delay from that due to a finite eccentricity. 
For a system as wide as \nltt\ tidal effects are likely to be
negligible \citep{fl13,bqaw13}.    Further relativistic effects may be harder to
disentangle: orbital period decay $\dot P_B$ will be of a magnitude
$-1\,\mu{\rm s}\,{\rm yr}^{-1}$, compared with a period derivative from
the \citet{shklovskii70} effect\footnote{The \citet{shklovskii70}
  $\dot P_B$ (also known as secular acceleration) is particularly
  large in the case of \nltt\ because of its 
proximity and large space velocity \citep{kv09}.  If measured, it can
be used to derive a geometric
distance constraint \citep{bb96}.} of $+24\,\mu{\rm s}\,{\rm yr}^{-1}$
(current data do not strongly constrain $\dot P_B$ because of the
short baseline, with $\dot P_B=(-1900\pm900)\,\mu{\rm s}\,{\rm yr}^{-1}$).
Therefore, an accurate determination of the distance (whose uncertainties
currently dominate the uncertainty in the period derivative) will be
necessary before any relativistic $\dot P_B$ can be measured.

\begin{figure}
%\plotone{../nltt11748_eclipse_times.eps}
\plotone{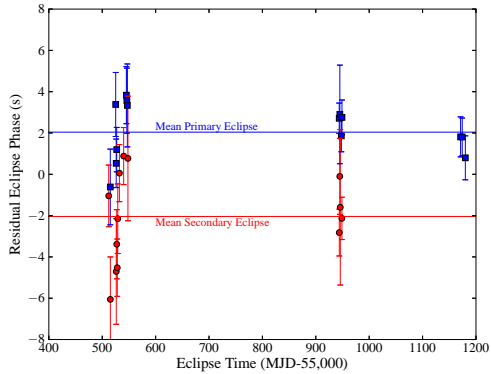}
\caption{Residual eclipse phase vs.\ eclipse time of \nltt, showing
  only the ULTRACAM measurements, although the measurements of
  \citet{sks+10} were included in the ephemeris calculations.  The
  primary eclipses are the blue squares, while the secondary eclipses
  are the red circles.  The data have been fit with a
  constant-frequency ephemeris.  We find an offset between the mean
  time of the primary eclipse compared with that of the secondary
  eclipse of $\Delta t=4.1\pm0.5$\,s.  }
\label{fig:dt}
\end{figure}

In comparison, the radial velocity constraints on the eccentricity are
considerably weaker: \citet{sks+10} determined $e<0.06$ (3$\sigma$),
while \citet{kapb+10b} and \citet{kvv10} both assumed circular orbits.
We refit all of the available radial velocity measurements from those
three papers.  Assuming a circular orbit we find
$K_1=273.3\pm0.4\,\kms$ (like \citealt{kapb+10b}, whose data dominate
the fit).  We also tried the eccentric orbit as parameterized by
\citet{dt91}.  The fit is consistent with a circular orbit, with
$e\cos\omega=0.001\pm0.004$ and $e\sin\omega=-0.004\pm0.004$, which
limits $e<0.017$ (3$\sigma$).
The eclipse durations can also weakly constrain the eccentricity, with
the ratio of the secondary eclipse to the primary eclipse duration
roughly given by $1+2e\sin \omega$ (for $e\ll 1$), although as noted
by \citet{winn11} this is typically less useful than the constraints
on $e\cos\omega$ from eclipse timing.  In the future we can fit for
this term directly.

While all data are satisfactorily fit by just a simple linear
ephemeris, we can also ask if a third body could be present in the
system.  Such a body, especially if on an inclined orbit, could
significantly speed up the merger of the inner binary and may alter
the evolution of the system \citep{thompson11}.
Any putative tertiary would likely be in a more distant circumbinary orbit,
since the interactions necessary to produce the ELM WD would have
disrupted closer companions. A tertiary would produce transit
timing variations \citep{hm05,assc05}, moving the eclipse times we
measure.  A full analysis of transit timing variations, including
non-linear orbital interactions, is beyond the
scope of this paper.  Instead, we did a limited analysis where we
considered the system to be sufficiently hierarchical such that the inner
binary was unperturbed (consistent with our measurements) and only its
center of mass moved due to the presence of the tertiary.  We took the
ephemeris residuals and fit a variety of periodic models, determining
for each trial period what the maximum amplitude could be
(marginalizing over phase) such that $\chi^2$ increased by 1 from the
linear ephemeris fit (adding additional terms in general decreases
$\chi^2$, but we wanted to see what the maximum possible amplitude
could be).  We found that for periods of 1--300\,days, the limit on any
sinusoidal component was $\lesssim 1\,$s (consistent with the rms
discussed above) or smaller than the orbit of
the inner binary.  Therefore, unless it is \textit{highly} inclined,
no tertiary with such a period is possible.  As we get to periods that
are longer than 300\,days, we no longer have sufficient data to
constrain a periodic signal, but here the constraints from our
polynomial fit also exclude any stellar-mass companion (an amplitude
of 1\,s at a period of 300\,days would require a mass of
$0.002\,M_\odot$ if the outer orbit is also edge-on).

\subsection{Secondary Temperature and Age Constraints}
\label{sec:temp}
As with the NIRI data, we can separate the fitting of the secondary
eclipse depth from the rest of the eclipse fitting and derive the
eclipse depths as a function of wavelength, $d_2(\lambda)$ (where we
also separate the 2010 and 2012 ULTRACAM observations).  Given the 8
secondary eclipse depths that we measure, we can determine the
ratio of the radius of the secondary to the primary $R_2/R_1$ as well
as the temperature of the secondary, $T_2$, given measurements of
$T_1$.  For that, we use the determination by \citet{kapb+10b}---$T_1=8690\pm140\,$K---largely consistent with the value determined
by \citet{kvv10} of $8580\pm50\,$K.  The secondary eclipse depths are
related to the wavelength-dependent flux ratios:
\beq
f(\lambda) \equiv \frac{F_{2}(\lambda)}{F_1(\lambda)} =
\frac{d_2(\lambda)}{1-d_2(\lambda)} = \frac{R_2^2 10^{-m(\lambda,T_2)/2.5}}{R_1^2 10^{-m(\lambda,T_1)/2.5}},
\label{eqn:flux}
\eeq
where $m(\lambda,T)$ is the absolute magnitude of a fiducial WD
with a temperature $T$ at wavelength $\lambda$.

To determine the secondary's temperature from the eclipse depths, we
first sample $T_1$ from the distribution ${\cal N}(8690,140)$ 1000
times.  Then, for each sample, we interpolate the synthetic photometry
of \citet[][using the $0.2\,M_\odot$ model]{tbg11} to determine the
photometry for the primary, $m(\lambda,T_1)$.  We then solve for the
temperature of the secondary and the radius ratio by minimizing the
$\chi^2$ statistic, comparing our measured $f(\lambda)$ with the
synthetic values (using the $0.7\,M_\odot$ grid for the secondary; the
results did not change if we used the $0.6\,M_\odot$ or $0.8\,M_\odot$
grids instead).

The resulting distribution of $T_2$ and $(R_2/R_1)$ is independent of
any assumptions in the global eclipse fitting.  We find
$T_2=7643\pm94\,$K and $R_2/R_1=0.255\pm0.002$.  However, the fits had
an average $\chi^2=12.6$ for 5 dof, mostly coming from
a small mismatch between the inferred eclipse depth at $g^\prime$
measured in 2010 versus 2012.  If we increase the uncertainties to have
a reduced $\chi^2=1$, then we find $T_2=7643\pm150\,$K and
$R_2/R_1=0.255\pm0.003$.  Note that the radius ratio here is fully
consistent with that inferred from the fit to the rest of the eclipse
shape (\Sref{fit} and \Tref{fullfitr2}).

We show the results in \Fref{depth} (including the results of the NIRI
data analysis), along with the results using
blackbodies for the flux distributions instead of the synthetic
photometry (as had been done by \citealt{sks+10} and others).  It is
apparent that the blackbody does not agree well and using the
synthetic photometry is vital.

The constraints using the separate wavelength-dependent eclipse depths
ended up being slightly less precise than the value from the full
eclipse fitting, although the two constraints are entirely consistent.
We therefore choose the values from \Tref{fullfitr2}, where we
determined a secondary temperature of $T_2=7600\pm120\,$K.  Using the
thin (thick) hydrogen atmosphere models for a 0.7\,$M_\odot$ CO WD, we
find a secondary age $\tau_2=1.70\pm0.09\,$Gyr ($1.58\pm0.07\,$Gyr) by
interpolating the cooling curves from \citet{tbg11}.  So, the envelope
uncertainty does not contribute significantly to the uncertainty in
the age of the secondary directly.  A bigger contribution is through
changes in the secondary mass $M_2$.

\subsection{Primary Radius and Distance Constraints}
Based on the measured $J$-band photometry \citep[$J=15.84\pm0.08$
  from][]{2mass}\footnote{\citet{kv09} incorrectly give the 2MASS
  $J$-band magnitude as $15.873\pm0.077$, but the online database
  lists 2MASS~J03451680+1748091 as having $J=15.837\pm0.077$.}, along
with an estimate for the extinction ($E(B-V)=0.10$ and $R_V=3.2$, from
\citealt{kapb+10b}), we can compare our measured radius with that
inferred from a parallax measurement ($\pi=5.6\pm0.9\,$mas; H.~Harris
2011, private communication).  We again use the \citet{tbg11} synthetic
photometry for the bolometric correction at the temperature determined
by \citet{kapb+10b}, and use $A_J=0.29A_V$.  Based on these data, we
infer $R_{1,\rm phot}=0.049\pm0.009\,R_\odot$.  This is fully
consistent (within 0.5\,$\sigma$) with our inferred values from the
eclipse fitting.  We can use this value along with the radius ratio
inferred from the eclipse shape (roughly $R_2/R_1=0.2567\pm0.0006$) to
determine $R_2=0.013\pm0.002\,R_\odot$.  From here, we can calculate $M_{2,{\rm thin}}=0.57\pm0.08\,M_{\odot}$ and $M_{2,{\rm
    thick}}=0.61\pm0.11\,M_{\odot}$, which makes use of evolutionary
models that were interpolated to the correct effective temperature
\citep{fbb01,blr01}.  These masses are a little lower than our
secondary masses calculated by the eclipse fitting, but differ by less
than $2\,\sigma$.  Inverting the problem, we infer based on our
eclipse fitting for $r_2=1.02$ a distance $d=159\pm8\,$pc
($\pi=6.3\pm0.3\,$mas), with the uncertainty dominated by the
uncertainty in the photometry.

Our analysis above included the effects of in-binary microlensing.
\citet{sks+10} assumed microlensing would modify the primary eclipse
depth, but did not show definitively that it was required for a good
fit \citep[cf.][]{mvs+13}. We fit the same data with the same
procedure as in \Sref{fit}, but did not allow for any decrease in the
depth of the primary eclipse from lensing.  The resulting fit was
adequate, although slightly worse than with lensing.  With no lensing
amplification, to match the depth of the primary eclipse requires a
reduced value of $R_2/R_1$ or a lower inclination.  We can do that by
increasing $R_1$ or decreasing $R_2$, but that is difficult as the
eclipse durations fix $R_1/a$ and $R_2/a$.  We end up accomplishing
this by increasing the masses, which widens the orbit (increasing $a$
to go along with the increase in $R_1$, and decreasing $R_2$ through
the mass-radius relation for the WD).  For $r_2=1.00$, we find
$M_1=0.16\,M_\odot$ and $M_2=0.74\,M_\odot$, along with
$R_1=0.044\,R_\odot$ ($\logg_1=6.36\pm0.04$).  While this combination
of parameters does not give as good a fit to the data as the fit with
lensing, the difference is not statistically significant, with an
increase in $\chi^2$ of 20 (the reduced $\chi^2$ increased from 1.018
to 1.019, for a chance of occurrence of about 50\%).  We
would need other independent information (such as a much more precise
parallax or a more precise time delay) to break the degeneracies.

\setlength{\tabcolsep}{0.02in} 
\begin{deluxetable*}{lrclcrclcrclcrcl}
\tablewidth{0pt}
\tabletypesize{\scriptsize}
\tablecaption{Eclipse Fitting Results\label{tab:fullfitr2}}
\tablehead{
\colhead{Quantity} & \multicolumn{3}{c}{Value: $r_2=1.00$} && 
\multicolumn{3}{c}{Value: $r_2=1.02$}&& \multicolumn{3}{c}{Value:
  $r_2=1.04$} && \multicolumn{3}{c}{Value: $r_2=1.06$}
}
\startdata
$M_1\tablenotemark{a}\,(M_\odot)$\dotfill & $0.136$& $\pm$& $0.007$ && $0.145$& $\pm$& $0.007$ && $0.153$& $\pm$& $0.007$ && $0.162$& $\pm$& $0.007$ \\
$R_1\tablenotemark{a}\,(R_\odot)$\dotfill & $0.0423$& $\pm$& $0.0004$ && $0.0426$& $\pm$& $0.0004$ && $0.0429$& $\pm$& $0.0004$ && $0.0433$& $\pm$& $0.0004$ \\
$K_1$\tablenotemark{a,b}\,(km\,s$^{-1}$)\dotfill & $273.4$& $\pm$& $0.5$ && $273.4$& $\pm$& $0.5$ && $273.4$& $\pm$& $0.5$ && $273.4$& $\pm$& $0.5$ \\
$M_2\,(M_\odot)$\dotfill & $0.707$& $\pm$& $0.008$ && $0.718$& $\pm$& $0.008$ && $0.729$& $\pm$& $0.008$ && $0.740$& $\pm$& $0.008$ \\
$R_2\,(R_\odot)$\dotfill & $0.0108$& $\pm$& $0.0001$ && $0.0109$& $\pm$& $0.0001$ && $0.0110$& $\pm$& $0.0001$ && $0.0111$& $\pm$& $0.0001$ \\
$i\tablenotemark{a}\,(\rm deg)$\dotfill & $89.67$ & $\pm$ & $0.12$ && $89.67$ & $\pm$ & $0.12$ && $89.66$ & $\pm$ & $0.12$ && $89.67$ & $\pm$ & $0.12$ \\
$a\,(R_\odot)$\dotfill & $1.514$ & $\pm$ & $0.009$ && $1.526$ & $\pm$ & $0.009$ && $1.538$ & $\pm$ & $0.009$ && $1.549$ & $\pm$ & $0.009$ \\
$q$\dotfill & $0.192$ &$\pm$ & $0.008$ && $0.201$ &$\pm$ & $0.008$ && $0.210$ &$\pm$ & $0.007$ && $0.219$ &$\pm$ & $0.007$ \\
$K_2$\,(km\,s$^{-1}$)\dotfill & $52.4$& $\pm$& $2.1$ && $55.0$& $\pm$& $2.1$ && $57.5$& $\pm$& $2.0$ && $60.0$& $\pm$& $2.0$ \\
$R_2/R_1$\dotfill & $0.2565$ & $\pm$ &  $0.0006$ && $0.2567$ & $\pm$ &  $0.0006$ && $0.2568$ & $\pm$ &  $0.0006$ && $0.2570$ & $\pm$ &  $0.0006$ \\
$\logg_1$\dotfill & $6.32$ & $\pm$ & $0.03$ && $6.34$ & $\pm$ & $0.03$ && $6.36$ & $\pm$ & $0.03$ && $6.38$ & $\pm$ & $0.03$ \\
$\logg_2$\dotfill & $8.22$ & $\pm$ & $0.01$ && $8.22$ & $\pm$ & $0.01$ && $8.22$ & $\pm$ & $0.01$ && $8.22$ & $\pm$ & $0.01$ \\
$T_1\tablenotemark{a,b}\,$(K)\dotfill & $8706$& $\pm$& $136$ && $8705$& $\pm$& $137$ && $8705$& $\pm$& $135$ && $8707$& $\pm$& $136$ \\
$T_2\tablenotemark{a}\,$(K)\dotfill & $7597$& $\pm$& $119$ && $7594$& $\pm$& $120$ && $7591$& $\pm$& $118$ && $7590$& $\pm$& $119$ \\
$t_0$\tablenotemark{a}\,(MBJD)\dotfill & 55772.041585 & $\pm$ & 0.000004&& 55772.041585 & $\pm$ & 0.000004&& 55772.041585 & $\pm$ & 0.000004&& 55772.041585 & $\pm$ & 0.000004\\
$P_B$\tablenotemark{a}\,(day)\dotfill & 0.235060485 & $\pm$ & 0.000000003&& 0.235060485 & $\pm$ & 0.000000003&& 0.235060485 & $\pm$ & 0.000000003&& 0.235060485 & $\pm$ & 0.000000003\\
$\Delta t$\tablenotemark{a}\,(s)\dotfill & 4.2 & $\pm$ & 0.6&& 4.2 & $\pm$ & 0.6&& 4.2 & $\pm$ & 0.6&& 4.2 & $\pm$ & 0.6\\
$\chi^2$/DOF\dotfill & $22978.8$ &/& 22566 && $22978.8$ &/& 22566 && $22978.7$ &/& 22566 && $22978.6$ &/& 22566 \\
\enddata
\tablenotetext{a}{Directly fit in the MCMC.  All other parameters are
  inferred.}
\tablenotetext{b}{Used a prior distribution based on spectroscopic
  observations.  All other prior distributions were flat.}
\end{deluxetable*}

\section{Conclusions}
Using extremely high-quality photometry combined with improved
modeling, we have determined the masses and radii of the WDs in the
\nltt\ binary to better than $\pm0.01\,M_\odot$ and
$\pm0.0005\,R_\odot$ statistical precision, although uncertainties in
the radius excess limit our final precision.  This analysis makes use of the
eclipse depth and shape, including corrections for gravitational
lensing, and is consistent with the R{\o}mer delay measured
independently from eclipse times, $\Delta t=4.2\pm0.6$\,s. This would
be the first detection of an observed R{\o}mer delay for ground-based
eclipse measurements (cf.\ \citealt{bmd+12,bwl12}), although in all of
these systems there is the possibility that the time delay is instead
related to a finite, but small, eccentricity.

Our mass measurement for the smallest plausible radius excess
($r_2=1.02$), $M_1=0.137\pm0.007\,M_\odot$, is significantly lower
than that inferred by \citet{kapb+10b} on the basis of the
\citet{pach07} evolutionary models or that inferred by \citet{ambc13}
from newer models.  Even for the highest value of $r_2$ that we
considered (1.06), we still find a primary mass of
$0.157\pm0.008\,M_\odot$, significantly below the $0.17-0.18\,M_\odot$
range discussed by \citet{kapb+10b} and \citet{ambc13}.  This may call
for a revision of those models to take into account the improved
observational constraints or it may indicate that an even higher
value of $r_2$ is more realistic.  In any case, our surface gravity is
lower than that from \citet{kapb+10b}, which was used by
\citet{ambc13}, while it is consistent to within $1\,\sigma$ with the
gravity measured by \citet{kvv10}.  With our \logg\ determination, the
mass is  closer to the prediction from \citet[][who find
  $M=0.174\,M_\odot$ for $\logg=6.40$ and $\log_{10}\Teff=3.93$,
  compared to $0.183\,M_\odot$ for the higher \logg]{ambc13}, although
again high values of $r_2$ are required.  However, our cooling age for
the secondary is a factor of 2--3 smaller than the cooling age of
$>4\,$Gyr for the
primary predicted by the \citet{ambc13} models for the lower mass.  This might be a reflection of
the non-monotonic evolution experienced by some ELM WDs, even though
\nltt\ seems to have a low enough mass that it would cool in a simpler
manner.  Further constraints on the mass
ratio from eclipse timing or direct detection of the secondary's
spectrum (the inferred values of $K_2$ in \Tref{fullfitr2} vary
significantly) could help resolve this question.

\acknowledgements We thank the anonymous referee for helpful comments.
This work was supported by the National Science Foundation under
grants PHY 11-25915 and AST 11-09174.  TRM was supported under a grant
from the UK's Science and Technology Facilities Council (STFC),
ST/F002599/1.  VSD, SPL and ULTRACAM were supported by the STFC.  ANW
was supported by the University of Wisconsin, Milwaukee Office of
Undergraduate Research. We thank the staff of Gemini for assisting in
planning and executing these demanding observations.  We thank
D.~Foreman-Mackey for help in using \texttt{emcee} and L.~Althaus for
supplying evolutionary models. Based on
observations obtained at the Gemini Observatory, which is operated by
the Association of Universities for Research in Astronomy, Inc., under
a cooperative agreement with the NSF on behalf of the Gemini
partnership: the National Science Foundation (United States), the
National Research Council (Canada), CONICYT (Chile), the Australian
Research Council (Australia), Minist\'{e}rio da Ci\^{e}ncia,
Tecnologia e Inova\c{c}\~{a}o (Brazil) and Ministerio de Ciencia,
Tecnolog\'{i}a e Innovaci\'{o}n Productiva (Argentina).

{\it Facilities:} \facility{Gemini:Gillett (NIRI)}, \facility{NTT
  (ULTRACAM)}, \facility{ING:Herschel (ULTRACAM)}

\bibliographystyle{apj} 
%\bibliography{wd}

\begin{thebibliography}{}

\bibitem[{Agol}(2002){Agol}]{agol02}
{Agol}, E. 2002, \apj, 579, 430

\bibitem[{Agol} {et~al.}(2005){Agol}, {Steffen}, {Sari}, \&  {Clarkson}]{assc05}
{Agol}, E., {Steffen}, J., {Sari}, R., \& {Clarkson}, W. 2005, \mnras, 359, 567

\bibitem[{Althaus} {et~al.}(2013){Althaus}, {Miller Bertolami}, \&  {C{\'o}rsico}]{ambc13}
{Althaus}, L.~G., {Miller Bertolami}, M.~M., \& {C{\'o}rsico},
A.~H. 2013,  \aap, 557, A19

\bibitem[{Antoniadis} {et~al.}(2012){Antoniadis}, {van Kerkwijk}, {Koester},  {Freire}, {Wex}, {Tauris}, {Kramer}, \& {Bassa}]{avkk+12}
{Antoniadis}, J., {van Kerkwijk}, M.~H., {Koester}, D., {Freire}, P.~C.~C.,  {Wex}, N., {Tauris}, T.~M., {Kramer}, M., \& {Bassa}, C.~G. 2012, \mnras,  423, 3316

\bibitem[{Barlow} {et~al.}(2012){Barlow}, {Wade}, \& {Liss}]{bwl12}
{Barlow}, B.~N., {Wade}, R.~A., \& {Liss}, S.~E. 2012, \apj, 753, 101

\bibitem[{Bassa} {et~al.}(2006){Bassa}, {van Kerkwijk}, {Koester}, \&  {Verbunt}]{bvkkv06}
{Bassa}, C.~G., {van Kerkwijk}, M.~H., {Koester}, D., \& {Verbunt}, F. 2006,  \aap, 456, 295

\bibitem[{Bell} \& {Bailes}(1996){Bell} \& {Bailes}]{bb96}
{Bell}, J.~F. \& {Bailes}, M. 1996, \apjl, 456, L33

\bibitem[{Bergeron} {et~al.}(2001){Bergeron}, {Leggett}, \& {Ruiz}]{blr01}
{Bergeron}, P., {Leggett}, S.~K., \& {Ruiz}, M.~T. 2001, \apjs, 133, 413

\bibitem[{Bergeron} {et~al.}(1992){Bergeron}, {Saffer}, \& {Liebert}]{bsl92}
{Bergeron}, P., {Saffer}, R.~A., \& {Liebert}, J. 1992, \apj, 394, 228

\bibitem[{Blandford} \& {Teukolsky}(1976){Blandford} \& {Teukolsky}]{bt76}
{Blandford}, R. \& {Teukolsky}, S.~A. 1976, \apj, 205, 580

\bibitem[{Bloemen} {et~al.}(2012){Bloemen}, {Marsh}, {Degroote},  {{\O}stensen}, {P{\'a}pics}, {Aerts}, {Koester}, {G{\"a}nsicke}, {Breedt},  {Lombaert}, {Pyrzas}, {Copperwheat}, {Exter}, {Raskin}, {Van Winckel},  {Prins}, {Pessemier}, {Fr{\'e}mat}, {Hensberge}, {Jorissen}, \& {Van  Eck}]{bmd+12}
{Bloemen}, S., {et al.} 2012,  \mnras, 422, 2600

\bibitem[{Brown} {et~al.}(2013){Brown}, {Kilic}, {Allende Prieto},  {Gianninas}, \& {Kenyon}]{bkap+13}
{Brown}, W.~R., {Kilic}, M., {Allende Prieto}, C., {Gianninas}, A., \&  {Kenyon}, S.~J. 2013, \apj, 769, 66

\bibitem[{Brown} {et~al.}(2011){Brown}, {Kilic}, {Hermes}, {Allende Prieto},  {Kenyon}, \& {Winget}]{bkh+11}
{Brown}, W.~R., {Kilic}, M., {Hermes}, J.~J., {Allende Prieto}, C., {Kenyon},  S.~J., \& {Winget}, D.~E. 2011, \apjl, 737, L23

\bibitem[{Burkart} {et~al.}(2013){Burkart}, {Quataert}, {Arras}, \&  {Weinberg}]{bqaw13}
{Burkart}, J., {Quataert}, E., {Arras}, P., \& {Weinberg}, N.~N. 2013,
\mnras, 433, 332

\bibitem[{Carter} \& {Winn}(2009){Carter} \& {Winn}]{cw09}
{Carter}, J.~A. \& {Winn}, J.~N. 2009, \apj, 704, 51

\bibitem[{Claret}(2000){Claret}]{claret00}
{Claret}, A. 2000, \aap, 363, 1081

\bibitem[{Damour} \& {Taylor}(1991){Damour} \& {Taylor}]{dt91}
{Damour}, T. \& {Taylor}, J.~H. 1991, \apj, 366, 501

\bibitem[{D'Antona} {et~al.}(2006){D'Antona}, {Ventura}, {Burderi}, \&  {Teodorescu}]{davb+06}
{D'Antona}, F., {Ventura}, P., {Burderi}, L., \& {Teodorescu}, A. 2006, \apj,  653, 1429

\bibitem[{Deloye} {et~al.}(2005){Deloye}, {Bildsten}, \& {Nelemans}]{dbn05}
{Deloye}, C.~J., {Bildsten}, L., \& {Nelemans}, G. 2005, \apj, 624, 934

\bibitem[{Dhillon} {et~al.}(2007){Dhillon}, {Marsh}, {Stevenson}, {Atkinson},  {Kerry}, {Peacocke}, {Vick}, {Beard}, {Ives}, {Lunney}, {McLay}, {Tierney},  {Kelly}, {Littlefair}, {Nicholson}, {Pashley}, {Harlaftis}, \&  {O'Brien}]{dms+07}
{Dhillon}, V.~S., {et al.} 2007, \mnras, 378, 825

\bibitem[{Fontaine} {et~al.}(2001){Fontaine}, {Brassard}, \&  {Bergeron}]{fbb01}
{Fontaine}, G., {Brassard}, P., \& {Bergeron}, P. 2001, \pasp, 113, 409

\bibitem[{Foreman-Mackey} {et~al.}(2013){Foreman-Mackey}, {Hogg}, {Lang}, \&  {Goodman}]{fmhlg13}
{Foreman-Mackey}, D., {Hogg}, D.~W., {Lang}, D., \& {Goodman}, J. 2013, \pasp,  125, 306

\bibitem[{Fuller} \& {Lai}(2013){Fuller} \& {Lai}]{fl13}
{Fuller}, J. \& {Lai}, D. 2013, \mnras, 430, 274

\bibitem[{Gianninas} {et~al.}(2013){Gianninas}, {Strickland}, {Kilic}, \&  {Bergeron}]{gskb13}
{Gianninas}, A., {Strickland}, B.~D., {Kilic}, M., \& {Bergeron}, P. 2013,  \apj, 766, 3

\bibitem[{Goodman} \& {Weare}(2010){Goodman} \& {Weare}]{gw10}
{Goodman}, J. \& {Weare}, J. 2010, Comm. App. Math. Comp. Sci., 5, 65

\bibitem[{Hobbs} {et~al.}(2006){Hobbs}, {Edwards}, \& {Manchester}]{hem06}
{Hobbs}, G.~B., {Edwards}, R.~T., \& {Manchester}, R.~N. 2006, \mnras, 369, 655

\bibitem[{Hodapp} {et~al.}(2003){Hodapp}, {Jensen}, {Irwin}, {Yamada},  {Chung}, {Fletcher}, {Robertson}, {Hora}, {Simons}, {Mays}, {Nolan}, {Bec},  {Merrill}, \& {Fowler}]{hji+03}
{Hodapp}, K.~W., {et al.} 2003, \pasp, 115,  1388

\bibitem[{Holman} \& {Murray}(2005){Holman} \& {Murray}]{hm05}
{Holman}, M.~J. \& {Murray}, N.~W. 2005, Science, 307, 1288

\bibitem[{Iben} \& {Tutukov}(1984){Iben} \& {Tutukov}]{it84}
{Iben}, Jr., I. \& {Tutukov}, A.~V. 1984, \apjs, 54, 335

\bibitem[{Kaplan}(2010){Kaplan}]{kaplan10}
{Kaplan}, D.~L. 2010, \apjl, 717, L108

\bibitem[{Kaplan} {et~al.}(2013){Kaplan}, {Bhalerao}, {van Kerkwijk},  {Koester}, {Kulkarni}, \& {Stovall}]{kbvk+13}
{Kaplan}, D.~L., {Bhalerao}, V.~B., {van Kerkwijk}, M.~H., {Koester}, D.,  {Kulkarni}, S.~R., \& {Stovall}, K. 2013, \apj, 765, 158

\bibitem[{Kaplan} {et~al.}(2012){Kaplan}, {Bildsten}, \& {Steinfadt}]{kbs12}
{Kaplan}, D.~L., {Bildsten}, L., \& {Steinfadt}, J.~D.~R. 2012, \apj, 758, 64

\bibitem[{Kawka} \& {Vennes}(2009){Kawka} \& {Vennes}]{kv09}
{Kawka}, A. \& {Vennes}, S. 2009, \aap, 506, L25

\bibitem[{Kawka} {et~al.}(2010){Kawka}, {Vennes}, \& {Vaccaro}]{kvv10}
{Kawka}, A., {Vennes}, S., \& {Vaccaro}, T.~R. 2010, \aap, 516, L7

\bibitem[{Kilic} {et~al.}(2010){Kilic}, {Allende Prieto}, {Brown},  {Ag{\"u}eros}, {Kenyon}, \& {Camilo}]{kapb+10b}
{Kilic}, M., {Allende Prieto}, C., {Brown}, W.~R., {Ag{\"u}eros}, M.~A.,  {Kenyon}, S.~J., \& {Camilo}, F. 2010, \apjl, 721, L158

\bibitem[{Kilic} {et~al.}(2012){Kilic}, {Brown}, {Allende Prieto}, {Kenyon},  {Heinke}, {Ag{\"u}eros}, \& {Kleinman}]{kbap+12}
{Kilic}, M., {Brown}, W.~R., {Allende Prieto}, C., {Kenyon}, S.~J., {Heinke},  C.~O., {Ag{\"u}eros}, M.~A., \& {Kleinman}, S.~J. 2012, \apj, 751, 141

\bibitem[{Lorimer} {et~al.}(1995){Lorimer}, {Lyne}, {Festin}, \&  {Nicastro}]{llfn95}
{Lorimer}, D.~R., {Lyne}, A.~G., {Festin}, L., \& {Nicastro}, L. 1995, \nat,  376, 393

\bibitem[{Maeder}(1973){Maeder}]{maeder73}
{Maeder}, A. 1973, \aap, 26, 215

\bibitem[{Mandel} \& {Agol}(2002){Mandel} \& {Agol}]{ma02}
{Mandel}, K. \& {Agol}, E. 2002, \apjl, 580, L171

\bibitem[{Marsh}(2001){Marsh}]{marsh01}
{Marsh}, T.~R. 2001, \mnras, 324, 547

\bibitem[{Marsh} {et~al.}(1995){Marsh}, {Dhillon}, \& {Duck}]{mdd95}
{Marsh}, T.~R., {Dhillon}, V.~S., \& {Duck}, S.~R. 1995, \mnras, 275, 828

\bibitem[{Marsh} {et~al.}(2004){Marsh}, {Nelemans}, \& {Steeghs}]{mns04}
{Marsh}, T.~R., {Nelemans}, G., \& {Steeghs}, D. 2004, \mnras, 350, 113

\bibitem[{Moffat}(1969){Moffat}]{moffat69}
{Moffat}, A.~F.~J. 1969, \aap, 3, 455

\bibitem[{Muirhead} {et~al.}(2013){Muirhead}, {Vanderburg}, {Shporer},  {Becker}, {Swift}, {Lloyd}, {Fuller}, {Zhao}, {Hinkley}, {Pineda}, {Bottom},  {Howard}, {von Braun}, {Boyajian}, {Law}, {Baranec}, {Riddle}, {Ramaprakash},  {Tendulkar}, {Bui}, {Burse}, {Chordia}, {Das}, {Dekany}, {Punnadi}, \&  {Johnson}]{mvs+13}
{Muirhead}, P.~S., {et al.} 2013, \apj, 767, 111

\bibitem[{Panei} {et~al.}(2007){Panei}, {Althaus}, {Chen}, \& {Han}]{pach07}
{Panei}, J.~A., {Althaus}, L.~G., {Chen}, X., \& {Han}, Z. 2007, \mnras, 382,  779

\bibitem[{Parsons} {et~al.}(2011){Parsons}, {Marsh}, {G{\"a}nsicke}, {Drake},  \& {Koester}]{pmg+11}
{Parsons}, S.~G., {Marsh}, T.~R., {G{\"a}nsicke}, B.~T., {Drake}, A.~J., \&  {Koester}, D. 2011, \apjl, 735, L30

\bibitem[{Paxton} {et~al.}(2011){Paxton}, {Bildsten}, {Dotter}, {Herwig},  {Lesaffre}, \& {Timmes}]{pbd+11}
{Paxton}, B., {Bildsten}, L., {Dotter}, A., {Herwig}, F., {Lesaffre}, P., \&  {Timmes}, F. 2011, \apjs, 192, 3

\bibitem[{Paxton} {et~al.}(2013){Paxton}, {Cantiello}, {Arras}, {Bildsten},  {Brown}, {Dotter}, {Mankovich}, {Montgomery}, {Stello}, {Timmes}, \&  {Townsend}]{pca+13}
{Paxton}, B., {et al.} 2013, \apjs, 208, 4

\bibitem[{Romero} {et~al.}(2012){Romero}, {C{\'o}rsico}, {Althaus}, {Kepler},  {Castanheira}, \& {Miller Bertolami}]{rca+12}
{Romero}, A.~D., {C{\'o}rsico}, A.~H., {Althaus}, L.~G., {Kepler}, S.~O.,  {Castanheira}, B.~G., \& {Miller Bertolami}, M.~M. 2012, \mnras, 420, 1462

\bibitem[{Serenelli} {et~al.}(2002){Serenelli}, {Althaus}, {Rohrmann}, \&  {Benvenuto}]{sarb02}
{Serenelli}, A.~M., {Althaus}, L.~G., {Rohrmann}, R.~D., \& {Benvenuto}, O.~G.  2002, \mnras, 337, 1091

\bibitem[{Shklovskii}(1970){Shklovskii}]{shklovskii70}
{Shklovskii}, I.~S. 1970, Soviet Astronomy, 13, 562

\bibitem[{Shporer} {et~al.}(2010){Shporer}, {Kaplan}, {Steinfadt}, {Bildsten},  {Howell}, \& {Mazeh}]{sks+10b}
{Shporer}, A., {Kaplan}, D.~L., {Steinfadt}, J.~D.~R., {Bildsten}, L.,  {Howell}, S.~B., \& {Mazeh}, T. 2010, \apjl, 725, L200

\bibitem[{Skrutskie} {et~al.}(2006){Skrutskie}, {Cutri}, {Stiening},  {Weinberg}, {Schneider}, {Carpenter}, {Beichman}, {Capps}, {Chester},  {Elias}, {Huchra}, {Liebert}, {Lonsdale}, {Monet}, {Price}, {Seitzer},  {Jarrett}, {Kirkpatrick}, {Gizis}, {Howard}, {Evans}, {Fowler}, {Fullmer},  {Hurt}, {Light}, {Kopan}, {Marsh}, {McCallon}, {Tam}, {Van Dyk}, \&  {Wheelock}]{2mass}
{Skrutskie}, M.~F., {et al.} 2006, \aj, 131, 1163

\bibitem[{Steinfadt} {et~al.}(2010){Steinfadt}, {Kaplan}, {Shporer},  {Bildsten}, \& {Howell}]{sks+10}
{Steinfadt}, J.~D.~R., {Kaplan}, D.~L., {Shporer}, A., {Bildsten}, L., \&  {Howell}, S.~B. 2010, \apjl, 716, L146

\bibitem[{Tauris} {et~al.}(2012){Tauris}, {Langer}, \& {Kramer}]{tlk12}
{Tauris}, T.~M., {Langer}, N., \& {Kramer}, M. 2012, \mnras, 425, 1601

\bibitem[{Thompson}(2011){Thompson}]{thompson11}
{Thompson}, T.~A. 2011, \apj, 741, 82

\bibitem[{Tremblay} \& {Bergeron}(2008){Tremblay} \& {Bergeron}]{tb08}
{Tremblay}, P.-E. \& {Bergeron}, P. 2008, \apj, 672, 1144

\bibitem[{Tremblay} {et~al.}(2011){Tremblay}, {Bergeron}, \&  {Gianninas}]{tbg11}
{Tremblay}, P.-E., {Bergeron}, P., \& {Gianninas}, A. 2011, \apj, 730, 128

\bibitem[{van Kerkwijk} {et~al.}(2005){van Kerkwijk}, {Bassa}, {Jacoby}, \&  {Jonker}]{vkbjj05}
{van Kerkwijk}, M.~H., {Bassa}, C.~G., {Jacoby}, B.~A., \& {Jonker}, P.~G.  2005, in ASP Conf. Ser., Vol. 328, Binary Radio Pulsars, ed. {F.~A.~Rasio \&  I.~H.~Stairs} (San Fransisco, CA: ASP), 357, arXiv:astro-ph/0405283

\bibitem[{Vennes} {et~al.}(2011){Vennes}, {Thorstensen}, {Kawka},  {N{\'e}meth}, {Skinner}, {Pigulski}, {St{\c e}{\' s}licki},  {Ko{\l}aczkowski}, \& {{\'S}r{\'o}dka}]{vtk+11}
{Vennes}, S., {et al.} 2011, \apjl, 737, L16

\bibitem[{Webbink}(1984){Webbink}]{webbink84}
{Webbink}, R.~F. 1984, \apj, 277, 355

\bibitem[{Winn}(2011){Winn}]{winn11}
{Winn}, J.~N. 2011, {Exoplanet Transits and Occultations}, ed. S.~{Piper},  55--77, arXiv:1001.2010

\end{thebibliography}

%% --------------------------------------------------------------------
%% Fri Oct 11 09:27:12 2013
%%   This file was generated automagically from the files
%%   nltt11748_ultracam.bbl and nltt11748_ultracam.tex using
%%     /Users/dlk//perl/nat2jour.pl
%%   This file should accompany nltt11748_ultracam-aas.tex.
%% --------------------------------------------------------------------

\end{document}